\documentclass[sigconf,screen,authorversion,nonacm,review=false,timestamp=false]{acmart}

\acmConference[ICPP Workshops’22]{51st International Conference on Parallel Processing Workshop}{August
29 -- September 1}{Bordeaux, France}

\usepackage{amsmath}
\usepackage{graphicx}
\usepackage{listings}
\usepackage[caption=true,font=footnotesize]{subfig}
\usepackage{verbatim}
\usepackage{xcolor}

\usepackage{tikz}              
\usepackage[export]{adjustbox} 
\usepackage{placeins}          

\definecolor{mauve}{rgb}{0.58,0,0.82}
\definecolor{mygreen}{rgb}{0,0.5,0}
\definecolor{mygray}{rgb}{0.4,0.4,0.4}
\definecolor{mymauve}{rgb}{0.58,0,0.82}
\definecolor{listinggray}{gray}{0.9}
\definecolor{lbcolor}{rgb}{0.95,0.95,0.95}

\lstdefinestyle{mystyle}{
	backgroundcolor=\color{lbcolor},
	tabsize=4,
	language=[GNU]C++,
	basicstyle= \footnotesize\ttfamily, 
	aboveskip={1.3\baselineskip},
	columns=fixed,
	showstringspaces=false,
	extendedchars=false,
	breaklines=true,
	frame=single,
	numbers=left,
	showtabs=false,
	showspaces=false,
	showstringspaces=false,
	identifierstyle=\ttfamily,
	keywordstyle=\color[rgb]{0,0,1},
	commentstyle=\color[rgb]{0,0.4,0},
	stringstyle=\color[rgb]{0.627,0.126,0.941},
	captionpos=b,
}

\AtBeginDocument{%
  \providecommand\BibTeX{{%
    \normalfont B\kern-0.5em{\scshape i\kern-0.25em b}\kern-0.8em\TeX}}}

\newcommand{\ompc}{OMPC}
\newcommand{\omp}{OpenMP}

\newcommand{\rsec}[1]{Section~\ref{sec:#1}}

\newcommand{\rfig}[1]{Figure~\ref{fig:#1}}

\newcommand{\rlst}[1]{Listing~\ref{lst:#1}}

\newcommand{\ttt}[1]{{\texttt{#1}}}
\newcommand{\tit}[1]{{\textit{#1}}}

\newcommand*\circled[1]{{\footnotesize\tikz[baseline=(char.base)]{
\node[shape=circle,draw,inner sep=1pt] (char) {#1};}}}

\newif\ifComments

\ifComments
\newcommand{\herve}[1]{\noindent\textcolor{red}{Herv\'{e}: {#1}}}
\newcommand{\guido}[1]{\noindent\textcolor{magenta}{Guido: {#1}}}
\newcommand{\marcio}[1]{\noindent\textcolor{green}{Marcio: {#1}}}
\newcommand{\gustavo}[1]{\noindent\textcolor{blue}{Gustavo: {#1}}}
\newcommand{\guilherme}[1]{\noindent\textcolor{orange}{Guilherme: {#1}}}
\newcommand{\del}[1]{\noindent\textcolor{gray}{Removed: {#1}}}

\newcommand{\short}[1]{\noindent\textcolor{blue}{ {#1}}}
\else
\newcommand{\herve}[1]{}
\newcommand{\guido}[1]{}
\newcommand{\marcio}[1]{}
\newcommand{\gustavo}[1]{}
\newcommand{\guilherme}[1]{}
\newcommand{\del}[1]{}

\newcommand{\short}[1]{}
\fi

\newcommand\blfootnote[1]{%
  \begingroup
  \renewcommand\thefootnote{}\footnote{#1}%
  \addtocounter{footnote}{-1}%
  \endgroup
}

\begin{document}

\title{The OpenMP Cluster Programming Model}


\author{Hervé Yviquel$^1$, Marcio Pereira$^1$, Emílio Francesquini$^2$, Guilherme Valarini$^1$, Gustavo Leite$^1$, \\ Pedro Rosso$^1$, Rodrigo Ceccato$^1$, Carla Cusihualpa$^1$, Vitoria Dias$^1$, Sandro Rigo$^1$, Alan Souza$^3$, \\ and Guido Araujo$^1$}
\affiliation{\institution{
    $^1$Institute of Computing -- University of Campinas -- UNICAMP -- Brazil \\
    $^2$Federal University of ABC -- UFABC -- Brazil \\
    $^3$Petrobras -- Brazil } \country{}}

\renewcommand{\shortauthors}{H. Yviquel et al.}




\begin{abstract}

Despite the various research initiatives and proposed programming models, efficient solutions for parallel programming in HPC clusters still rely on a complex combination of different programming models (e.g., OpenMP and MPI), languages (e.g., C++ and CUDA), and specialized runtimes (e.g., Charm++ and Legion). On the other hand, task parallelism has shown to be an efficient and seamless programming model for clusters. This paper introduces OpenMP Cluster (OMPC), a task-parallel model that extends OpenMP for cluster programming. OMPC leverages OpenMP's offloading standard to distribute annotated regions of code across the nodes of a distributed system. To achieve that it hides MPI-based data distribution and load-balancing mechanisms behind OpenMP task dependencies. Given its compliance with OpenMP, OMPC  allows applications to use the same programming model to exploit intra- and inter-node parallelism, thus simplifying the development process and maintenance. We evaluated OMPC using Task Bench, a synthetic benchmark focused on task parallelism, comparing its performance against other distributed runtimes. Experimental results show that OMPC can deliver up to 1.53x and 2.43x better performance than Charm++ on CCR and scalability experiments, respectively. Experiments also show that OMPC performance weakly scales for both Task Bench and a real-world seismic imaging application.

\end{abstract}

\maketitle

\blfootnote{Contact authors: hyviquel@unicamp.br and guido@unicamp.br}

\section{Introduction}

Limitations on the power-density of modern processor cores and the continuous growth in the utilization of data centers~\cite{Crago2015} and supercomputers~\cite{Meuer2014} have led to an increasing interest in the adoption of power-efficient hardware accelerators (e.g., GPU and FPGA) and novel instruction-set architecture extensions (e.g., IBM MMA and Intel AMX). Such heterogeneous architectures have considerably increased the complexity of programming parallel architectures, which already suffered from the combination of multiple programming models and languages. A modern scientific or machine learning application, for example, requires an entangled combination of message-based (e.g., MPI), multi-threading, and GPU programming (e.g., CUDA) to execute.

To simplify this task, directive-based programming models such as those found in OpenMP have become widely adopted, given their simplicity and ease of use. In OpenMP, program fragments (e.g., hot loops) are annotated by the programmer to expose parallelism. Recent versions of this standard~\cite{OpenMP2013} include new directives that enable task parallelism and the transfer of computation to acceleration devices using a technique called \textit{computation offloading} (or simply offloading). From the programmer's point of view, the program starts running on the host (e.g., CPU), and when a program snippet annotated with the OpenMP clause is reached, code and data are transferred to the given device for execution, returning the control flow to the host upon completion. While offloading has been widely used to move computation to GPUs using languages such as CUDA, or OpenMP, these approaches are, in most cases, restricted to a single computing node. Transparent offloading across a large cluster with hundreds or thousands of nodes is a much more complex and cumbersome task.

In this paper, we propose \ompc\footnote{OMPC is available at \url{https://ompcluster.gitlab.io/}}, an OpenMP task parallelism-based execution model for clusters. \ompc~allows for the offloading of complex scientific tasks across HPC cluster nodes in a transparent and balanced way. To achieve that we extended OpenMP with a new device plug-in that models a cluster computing node. This new device enables a set of powerful features that are hidden away from the programmer: (a) an underlying MPI communication layer for inter-node communication; (b) an event handler that transparently offloads tasks and data to cluster nodes; and (c) a cluster-wide HEFT-based task scheduler~\cite{Topcuoglu02} that balances the workload distribution across the cluster. These tasks are hidden behind the OpenMP programming model in such a way that a programmer sees computing tasks being offloaded to cluster nodes, similarly as he/she would see tasks being distributed across cores of a single machine in a regular \omp~program. Challenging the design of \ompc~was to ensure that the new model is still compliant with \omp. To ensure that, a new device plugin was designed around the \omp~original runtime, thus enabling programmers to still count on the second level of parallelism inside each cluster node. In fact, inside a node, any combination of OpenMP, CUDA, or MPI can still be used, as in a typical parallel HPC application. Overall, \ompc~leverages the simplicity of OpenMP to build a system that can enable programmers to use it to scale parallelism from a single node to a complex HPC cluster.

To have a glimpse of the improvements that \ompc~can bring to the performance of applications, please consider \rfig{speedup}, which shows \ompc~is able to deliver execution time for four \tit{Task Bench}~\cite{taskbench} kernels up to 1.53x and 2.43x faster when compared to Charm++, a well-known task-parallel programming model. The baseline of the figure is the same kernel but programmed using a synchronous MPI model on a cluster of up to 64 nodes with 24 cores each. The remaining of the paper provides additional details on how \ompc~achieves this speedup and makes a thorough comparative analysis of its performance on Task Bench benchmarks, while demonstrating its scalability for a real-world geophysics seismic application used to detect oil reservoirs.

The main contributions of this paper are the following:
\begin{itemize}
\item An approach for \tit{programming scalability}, in which a programmer can start writing a small prototype program for a single node, and gradually scale it out for a larger cluster with multiple nodes, all using the same OpenMP programming model.
\item A new device plugin that extends the \omp~Acceleration Model for a cluster node, implemented in such a way to not interfere with the regular \omp~runtime.
\item An event handling system, which hides the complexity of MPI inter-task communication behind OpenMP dependencies.
\end{itemize}

This paper is organized as follows. \rsec{background} details the task and offloading execution model available in the OpenMP standard. \rsec{ompc} introduces and gives an overview of the proposed \ompc~model. \rsec{runtime} gives a detailed account of the inner workings of \ompc, including its MPI event-based systems, data manager and scheduler. \rsec{relatedworks} discusses related works, and \rsec{experiments} the experiments realized to validate \ompc. Finally, \rsec{futurework} presents potential future works and \rsec{conclusion} concludes the paper.

\section{Background}
\label{sec:background}

This section introduces the concepts of computation offloading and task-based parallelism that are needed for a thorough understanding of the ideas proposed by the \ompc~programming model.

Task-based programming models (e.g., OpenSs and Charm++) have proven to be a scalable and flexible approach to extract both regular and irregular parallelism from applications~\cite{Duran2011,Kale1993,Bauer2012}. In this model, the programmer uses program constructs or code annotations to mark code regions in the program as tasks, which are later used by a runtime library to manage the parallel execution of task regions.

Tasks were introduced in OpenMP 3.0 to expose a higher degree of parallelism through the non-blocking execution of code fragments annotated with the \texttt{task} directives. The annotated code is outlined as tasks that are dispatched to the OpenMP runtime by the \textit{control thread}. Dependencies between tasks are specified by the \ttt{depend} clause, which is used to define the set of input variables that the task depends on, and the output variables that the task writes to. Whenever the dependencies of a given task are satisfied, the OpenMP runtime dispatches it to a \textit{ready queue} which feeds a pool of \textit{worker threads}. The OpenMP runtime manages the task graph, handles data management, and creates and synchronizes threads, among other activities.

The original OpenMP \texttt{task} directive was designed having a multicore architecture in mind. With the emergence of acceleration devices (e.g., GPUs and FPGAs) the standard was extended with the \textit{OpenMP accelerator model} (OpenMP 4.X) to enable computation offloading to such devices. The \texttt{target} directive was then introduced to have a similar semantics as the \texttt{task} directive, but for acceleration devices. It uses similar clauses (e.g., \texttt{depend}), but aims at offloading the computation to an accelerator instead of a CPU core. The \texttt{map} clause was introduced to specify the direction of the transfer operation that the host must execute with the given data. The most common transfer operations are \texttt{to}, \texttt{from}, and \texttt{tofrom}. Finally, clause \texttt{nowait} was proposed to state that the \texttt{target} is non-blocking and thus can run asynchronously on the accelerator.

To clarify this explanation, please consider the code fragment of \rlst{targettasks}. In that code, two tasks \ttt{foo} and \ttt{bar} are annotated with the \ttt{target} directive from \omp~that allows them to be offloaded for computation on accelerators. Under the \omp~specification~\cite{openmp52} this implies that the code and data for tasks \ttt{foo} and \ttt{bar} are dispatched to an accelerator (e.g., GPU) and the host machine, which runs the code in \rlst{targettasks}, will be responsible for moving their data around so as to satisfy all the dependencies between them.

\begin{figure*}
\begin{minipage}{0.7\linewidth}
\begin{lstlisting}[style=mystyle, label={lst:targettasks}, caption={OpenMP target tasks.}]
#pragma omp target enter data map(to: A[:N]) nowait depend(out: *A)
#pragma omp target nowait depend(inout: *A)
foo(A)
#pragma omp target nowait depend(inout: *A)
bar(A)
#pragma omp target exit data map(release: A[:N]) nowait depend(out: *A)
\end{lstlisting}
\end{minipage}
\end{figure*}

Now, let's analyze the semantics of this code. First, in line 1, the vector \ttt{A} is copied from the host memory to the accelerator memory, a task called \tit{offloading}. This is followed by lines 2-3 which dispatch \ttt{foo}'s code to the accelerator, execute it, storing its result on top of \ttt{A}, and finally bring back \ttt{A} to the host memory. The \ttt{depend(inout: *A)} clause is used to mark that \ttt{A} is read and written by \ttt{foo}. In lines 4-5 a similar computation happens, this time with the \ttt{bar} task. The \ttt{target} directive in line 4 assigns \ttt{bar} to an accelerator, and informs that it will read \ttt{A} (previously written by \ttt{foo}), which is stored in host memory. As before, the runtime will read \ttt{A} from the host memory, move it to the accelerator assigned to \ttt{bar}, execute \ttt{bar} storing the result on the accelerator memory, and bring \ttt{A} back again to host memory. Finally, line 6 ends the execution. As the reader might have noticed, all \ttt{target} directives have a \ttt{nowait} clause to state that \ttt{foo} and \ttt{bar} are executed asynchronously. Hence, the \omp~runtime ensures that all \ttt{depend} clauses in lines 2-4 are satisfied as defined by the programmer.

The user should remember that \omp~was designed to run on a single \textit{host} machine (i.e., node). Thus a set of cores from the host is assigned to run the code of \rlst{targettasks} and the \omp~runtime. The accelerators (e.g., GPUs) that run \ttt{foo} and \ttt{bar} are also connected to the same machine (typically via PCIe), and the communication between the tasks uses the accelerator and host memories. Also, if no accelerator exists in the machine, the \omp~runtime falls back the execution of \ttt{foo} and \ttt{bar} to regular \omp~ tasks, i.e., they are both executed by the host cores.

The OpenMP accelerator model has been used to offload computation to acceleration devices like GPUs~\cite{Huber2022}, Xeon Phi, and FPGAs architectures~\cite{Mayer2019}. In all such approaches the OpenMP runtime offloads data and code to a local (i.e., \tit{intra-node}) device, typically connected to the multicore host using a PCIex, NVLINK or CAPI interface. Not much research has been done on extending OpenMP so that computation offloading is performed across the nodes of a larger cluster, as proposed in this paper. The goal of this work is to provide programmers with an approach, based on a well-known programming model (\omp), that allows them to scale an application from a single machine prototype to a large computing cluster, without much effort.

\section{The \ompc~Programming Model}
\label{sec:ompc}

This section highlights the main differences between the proposed OpenMP model when compared to the OpenMP Accelerator Model. OMPC was designed to become a natural extension of the \omp~semantics to a cluster. To achieve that, a very simple abstraction was adopted, which extends the notion of \tit{cores} in \omp~to \tit{nodes} in \ompc. To better explain that, please consider the following three examples. First, while in \omp~a \ttt{target} assigns a task to an accelerator on the same machine, in \ompc~the task is assigned to a cluster node. Second, while in \omp~the \ttt{depend} clause moves data to/from host and accelerator memories (both on the same machine) in OMPC, the \ttt{depend} clause uses a set of underlying calls to an MPI subsystem to move data between nodes of the cluster. Overall, if the programmer keeps in mind that a core in \omp~corresponds to a node in \ompc, the semantics of the \omp~specification still hold for \ompc. For example, the \omp~code in \rlst{targettasks} does not need to be changed when it runs on a cluster over the \ompc~runtime. In this scenario, tasks \ttt{foo} and \ttt{bar} are assigned to cluster nodes and vector \ttt{A} is transferred between the nodes using the \ompc~implementation for the \ttt{depend} clause, which uses efficient MPI calls to move \ttt{A} from \ttt{foo} to \ttt{bar}.

\subsection{Execution model}
\label{sec:proposal}

\begin{figure}[!t]
	\centering
	\includegraphics[width=1\linewidth]{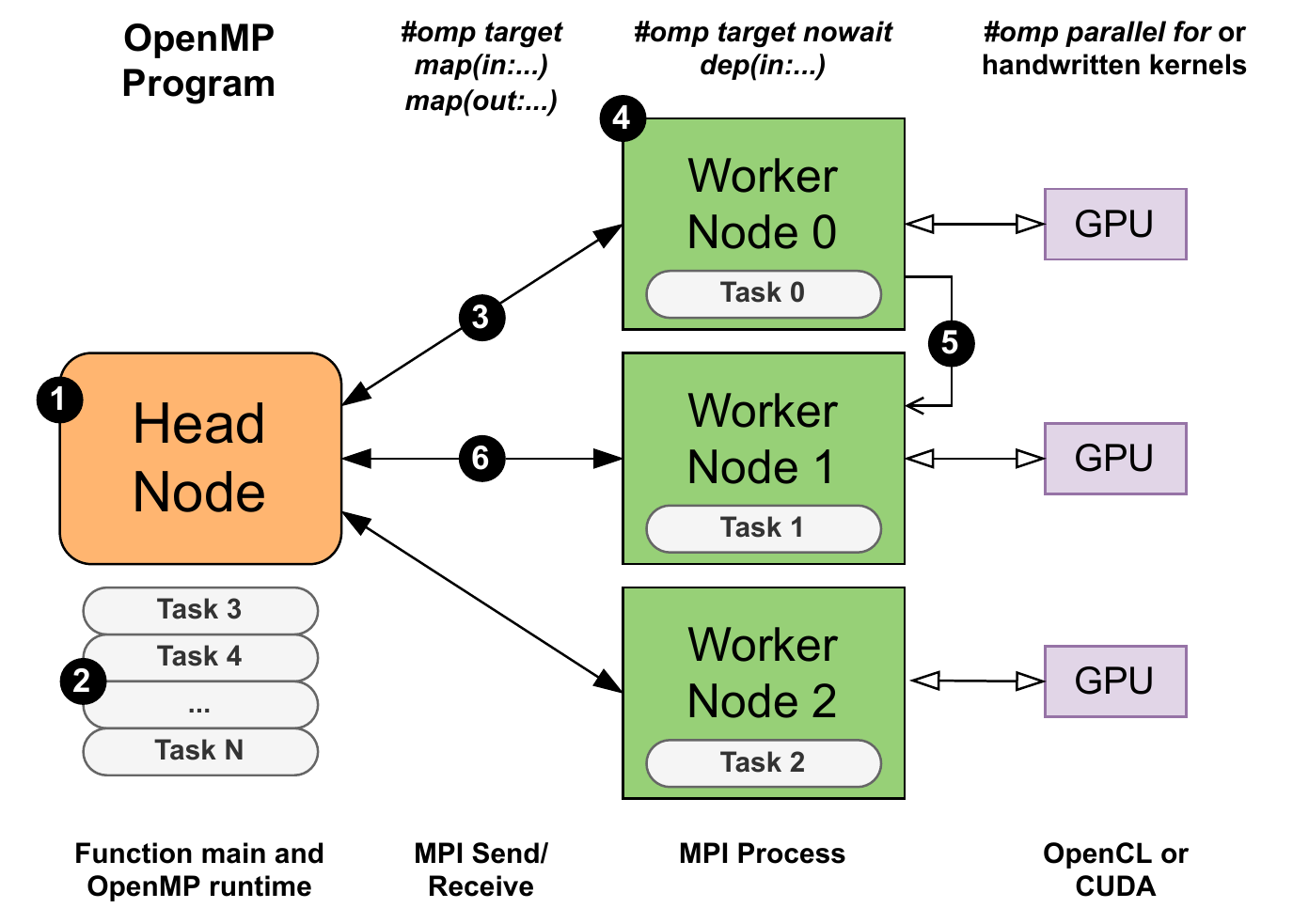}
	\caption{Execution Model of the Heterogeneous Cluster Device using OpenMP}
	\label{fig:flow}
\end{figure}

As shown in \rfig{flow}, a typical cluster consists of a \tit{head node} (which executes the \ompc~runtime), and a set of \tit{worker nodes}. The execution of the annotated code in the cluster has the following workflow. \circled{1} The user executes his/her program from the head node; \circled{2} When execution reaches the OpenMP kernel, the \omp~runtime automatically creates the tasks (without executing them) and adds them to a dedicated pool where they wait for execution; for the sake of clarity, assume that tasks \ttt{foo} and \ttt{bar} from \rlst{targettasks} correspond respectively to \ttt{Tasks} \ttt{0} and \ttt{1} of \rfig{flow}. \circled{3} The \ompc~runtime distributes the task (e.g., \ttt{foo}) and its input data (e.g., vector \ttt{A}) to be executed by some working nodes (e.g., \ttt{foo} to \ttt{Node 0}). To do that, it uses calls to the underlying MPI sub-system and follows a scheduling strategy (see~\rsec{heft}) to balance the computational workload. \circled{4} The worker node computes the received data. \circled{5} The \ompc~runtime forwards the result (i.e., the output of \ttt{foo}) to the task that depends on it (i.e.,~\ttt{bar} at~\ttt{Node 1}), also using MPI calls, removes the task from the dependency graph, and updates the dependencies. On the head node side, the tracking of the worker nodes will be done through the management of the dependency graph as described in the \omp~specification. \circled{6} To conclude the computation, the \ompc~runtime brings the vector \ttt{A} to the head node.

A relevant point must be highlighted regarding the code offloaded to the nodes. The codes of \ttt{foo} or \ttt{bar} are just some regular codes which could also benefit from the second level of parallelism. For example, if \ttt{foo} had a loop annotated with a \ttt{parallel for} directive, it still could benefit from \omp~parallelism inside the node. \ttt{Foo} or \ttt{bar} could also be written in OpenCL or CUDA, as a programmer would do in a distributed cluster. As said before, \ompc~was designed to have a smooth integration to the regular \omp~runtime.

\ompc~was also designed with fault tolerance in mind. To enable that, each node in \ompc~(head node and worker nodes) has a heartbeat mechanism, connected in a ring topology, which allows nodes to monitor their neighbors. Thus, if a node fails, the system detects and restarts the failed tasks. Fault tolerance work on \ompc~is underway and will be released in a future version. This mechanism will not be discussed further in this paper.

\section{The \ompc~Runtime}
\label{sec:runtime}


This section covers the main modules behind the OMPC Runtime. As presented in~\rsec{libomptarget}, it relies on a flexible implementation of the OpenMP accelerator model developed within the OpenMP offloading library~\cite{Antao2016} (known as \textit{libomptarget}) available within the LLVM project~\cite{Lattner}. The runtime includes three main components: (a) an MPI-based event system, described in~\rsec{event}; (b) an automatic Data Management module, described in~\rsec{datamanager}; and (c) a Task Scheduler based on the HEFT algorithm~\cite{Topcuoglu02}, described in~\rsec{heft}.

\subsection{OpenMP Target library integration}
\label{sec:libomptarget}

\begin{figure}
 	\centering
 	\includegraphics[width=1\linewidth]{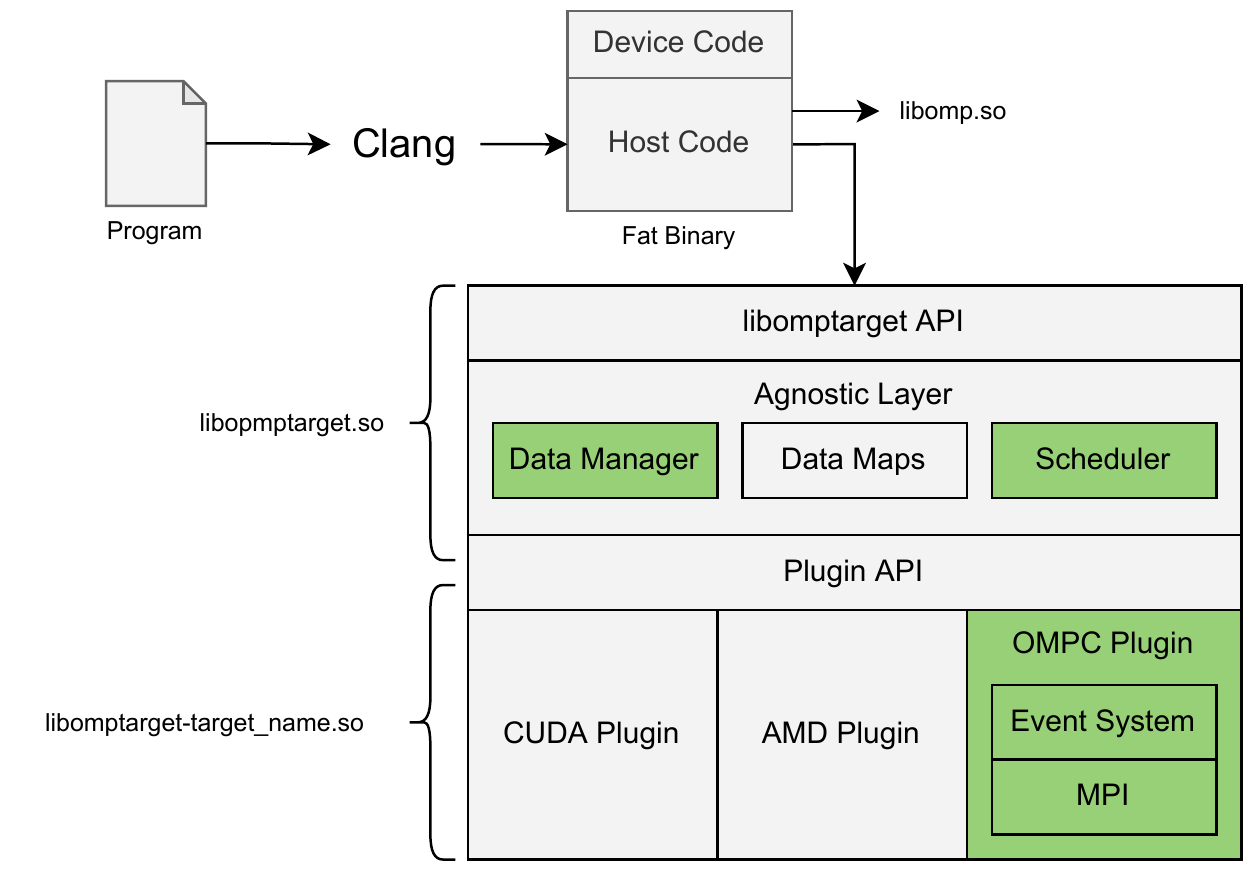}
    \caption{OMPC integration with LLVM's libomptarget library (OMPC contributions in green)}
 	\label{fig:libomptarget}
\end{figure}

LLVM's libomptarget implements a support library that, alongside the OpenMP runtime, allows the program to offload computation to accelerators. The library was implemented in a modular manner with extensibility in mind, as summarised in \rfig{libomptarget}. At the top layer of its design, libomptarget exposes a set of functions, used by the Clang code generator, that allows the program to interact with the target device at a high level. Next, the agnostic layer translates such function calls into low-level interactions with the accelerator (e.g., memory allocation, data movement, and code execution). Here, libomptarget also keeps track of where data is stored and which device should execute a target region. At the bottom layer, a streamlined "plugin" interface mediates the interactions between the agnostic layer and the target devices. Each device-specific plugin behaves as a driver for an accelerator, directly communicating and managing it. At this level, for example, one may encounter a plugin that uses the CUDA library to manage GPUs, or the OMPC plugin that relies on MPI calls to allow the program to run on a distributed environment.

Some changes were made to many of the described layers in order to correctly implement OMPC. We extended the Clang code generation and the OpenMP runtime to enrich the task graph information available to the libomptarget agnostic layer. The layer itself was also changed to allow \ompc~to manage and forward data between the multiple nodes (\rsec{datamanager}), and to automatically schedule the tasks over the cluster (\rsec{heft}). Finally, we implemented an OMPC plugin that allowed the head node to interact and command the many worker nodes using a multi-threaded MPI environment managed by the event system (\rsec{event}).

\subsection{MPI-based Event System}
\label{sec:event}


The usual data-parallel MPI applications are well known for their high complexity and strong coupling among the cluster nodes. All processes execute the same program and have matching communication primitive calls done in pre-defined rounds. A recent study shows that even if MPI applications are becoming more and more multi-threaded, most of them are using the funneled threading model~\cite{mpistudy}. Although OMPC uses MPI underneath, its behavior is far from similar to this usual case.  OMPC task parallelism nature is much more dynamic: different code regions with different purposes may execute at the same time on-demand, generating unrelated message exchanges between the cluster nodes and their many threads. Such OMPC characteristics make the usage of MPI as a communication engine challenging and difficult to implement correctly and in a performant manner, when even a single non-matched message may stall the whole program. To address this problem, OMPC integrates a distributed event system into the plugin that allows for a tighter and more manageable MPI usage while creating a safe environment for concurrent and dynamic MPI communications. \rfig{event-system-arch} shows all the elements that compose the event system described below:

\begin{figure*}[!ht]
 	\centering
 	\includegraphics[width=1\linewidth]{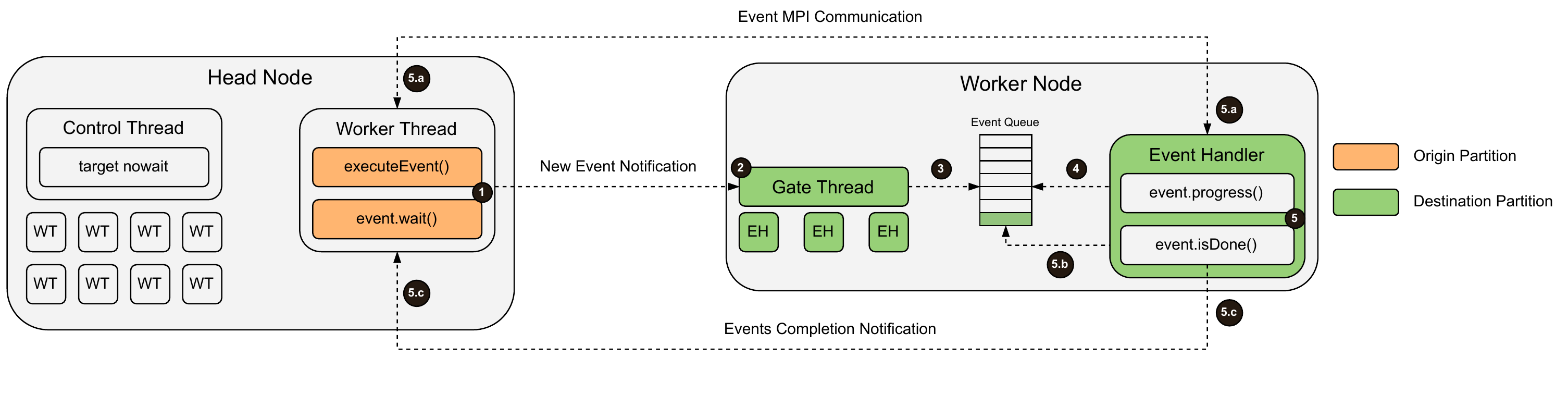}
    \caption{Distributed event system architecture and its inner workings}
 	\label{fig:event-system-arch}
\end{figure*}

\begin{description}
    \item[Events:] logical units that encapsulate multiple MPI messages representing different types of actions performed between two or more cluster nodes. Each event type specifies an origin and a destination (with the origin usually being the head node), ensuring that matching MPI calls are always done correctly between the processes. The actions that  can be currently performed are the allocation and removal of memory regions, the submission and retrieval of data, the indirect forwarding of data between two remote nodes (e.g., worker nodes), and the execution of a target region. These actions have a one-to-one match to all the required functions that a device plugin must implement in the libomptarget library.
    \item[Gate thread:] responsible for receiving new event notifications from other processes and enqueuing their destination counterpart into a local event queue.
    \item[Event handler:] a set of threads with the sole purpose of executing the events present in the local queue, and re-enqueuing them when any I/O operation is needed.
\end{description}

Events in the event system always flow in the same way as shown by \rfig{event-system-arch}. Let's say the control thread at the head node encounters a target region and dispatches it to a local worker thread, which will need to run an execution event. First, the worker thread creates the origin part of the event ($E_{O}$) and collects all of the arguments of the task~\circled{1}. Then it waits on the event execution, blocking the thread and starting the event itself. The event system sends a new event notification to the destination process~\circled{2}, which is received by its local \tit{gate thread}, creating the destination part of the event ($E_{D}$) and enqueueing it in the local queue~\circled{3}. Later, one of the local \tit{event handlers} acquires $E_{D}$ and starts its execution~\circled{5}. Here multiple messages may be exchanged with the head node~\circled{5a}, receiving which function must be executed and its parameters. Afterward, if an I/O is still pending, the event is re-enqueued~\circled{5b}, otherwise, an event completion notification~\circled{5c} is sent to the origin process, allowing for any thread waiting on the event to continue, thus completing the event flow.

Many messages may be exchanged during the execution of the event system, and thus great care must be taken to ensure that an exclusive "channel" is created between the communicating parties, so as to exclude any possibility of cross-talk between different events. By using MPI's message matching specification, each event receives a unique MPI tag local to the origin process which is shared with the destination process in the new event notification. This way, all MPI communications between the processes use the same tag, which, alongside the origin and destination ranks, ensures that only a given event will receive its own messages.

MPI implementations usually map different network hardware contexts to different MPI Communicators, with the latest MPI standard, even allowing such contexts to be mapped to different MPI tags~\cite{mpivcis}. Thus, to better utilize the hardware capabilities on the largest set of cluster configurations, the event system creates a set of Communicators at the beginning of the program. Whenever a new event is created, one communicator is selected in a round-robin fashion based on its MPI tag. This ensures that the events work in isolation, and allows for better utilization of the concurrent capabilities of the network hardware.

\subsection{Data Manager}
\label{sec:datamanager}


The OpenMP standard allows a program to register a buffer to be used by multiple target regions, sending the data only once to the offloading device and reusing it throughout the execution. This functionality allows reducing the communication between the host and the accelerator while improving the code clarity by reducing the number of map clauses in target regions. Although this feature can be easily implemented when using a single device (e.g., GPUs), it is not immediately compatible with OMPC, since it exposes multiple distributed address spaces over a single device. Thus, a Data Management (DM) module was integrated inside the runtime to maintain the data coherency over the cluster nodes. By analyzing the task dependencies between target regions, OMPC automatically forwards data between worker nodes without using the host (i.e., head node) as an intermediate location, dramatically improving performance.

The Data Management module is located at the Agnostic Layer (\rfig{libomptarget}). By intercepting the target clause executions, it keeps track of every buffer location by maintaining a set of data maps for each worker node. When OMPC executes a target region, DM checks those maps and chooses how to forward the needed data using the following criteria:

\begin{description}
    \item[Enter data clauses:] the task graph is analyzed after the task scheduling is completed and each buffer is sent to the first node that will use it.
    \item[Exit data clauses:] the DM consults its data maps and retrieves the buffer from any of its previous locations, sending it to the head node. If needed (i.e., the program will not use the data anymore), the buffer is removed from the entire cluster.
    \item[Target regions:] after the target task is scheduled to a specific node, the DM consults its data maps before executing the task. If a buffer is not already present in the node, it is forwarded (i.e., copied) from its most recent location (e.g., the head node or one of the worker nodes). After the task execution, if the buffer was marked as an \ttt{inout} dependency, the DM keeps the buffer only on the recent node, removing it from all previous locations. Otherwise (i.e., it is a read-only data), the buffer is kept in all of its previous locations, allowing it to be reused in the future.
\end{description}

As an example, consider the program in \rlst{targettasks} and \rfig{flow} as its execution flow. First, the entire task graph is built and scheduled \circled{2} (as explained in \rsec{heft}), with \ttt{foo} being mapped to worker node 0 (Task 0) and \ttt{bar} to worker node 1 (Task 1). Then, the DM intercepts the \ttt{enter data} clause execution \circled{1} and, by analyzing the task graph, it forwards \ttt{A} to the first node that will use it \circled{3}. Next, the program execution advances, running \ttt{foo} at the assigned node \circled{4}, which may update the buffer \ttt{A} (i.e., \ttt{inout} dependency). After \ttt{foo}'s completion, the program starts the execution of the task \ttt{bar} \circled{6}. Since \ttt{A} may be updated by \ttt{foo}, the DM intercepts the task execution and, instead of sending the buffer from the head node to the worker node 1, it copies it from the worker node 0 \circled{5}. After the Task 1 execution, the DM checks if \ttt{A} was updated by searching for an \ttt{inout} dependency on \ttt{bar}'s dependency list. If it is found, which is the case, the DM leaves the data present only on worker node 1, removing all outdated copies on the other worker nodes (i.e., worker node 0). This ensures that, if \ttt{A} is used again, the buffer must be acquired from the most up-to-date location. Finally, the program encounters the \ttt{target exit data} clause and removes \ttt{A} from the entire cluster.

The current implementation of the Data Management module has its own pros and cons. Such a mechanism allows the OMPC runtime to use information that would already be present in a task-based OpenMP program (e.g., dependencies list on tasks) to infer an intent to write to a target memory region, easing the adaptation of such programs into OMPC. The DM also allows read-only data to be distributed to the cluster on demand, which opens the possibility for one-to-many MPI optimizations based o the task graph in the future. But, it also limits how a program can be written, since a target task must always include every used data in its dependency list, hurting the code clarity and flexibility. Other possible mechanisms are listed in ~\rsec{futurework} and may replace the current one in the future.

\subsection{HEFT-based Task Scheduler}
\label{sec:heft}

In the \omp~programming model, tasks are created and executed inside a parallel region. While the control thread creates the tasks, the worker threads start executing them right away. LLVM's \omp~implementation uses a work-stealing algorithm to schedule tasks. Each thread has its own work queue and when their queue becomes empty, the threads proceed to steal tasks from their siblings. To ensure all tasks have been executed by the end of the parallel region, the \omp~standard requires the compiler to insert an implicit barrier immediately after the parallel region.

Work-stealing is a dynamic scheduling algorithm in the sense that it does not require the entire task graph to be available before it can start scheduling or executing tasks. This approach works well in multi-threaded applications where the cost of stealing a task between threads is small. However, employing such an algorithm in a multi-node application incurs transferring data over the network whenever one process steals a task from another, causing a prohibitively large overhead if done frequently. In contrast, static schedulers have more information available at the scheduled times and therefore are capable of mapping tasks to devices in such a way that reduces communication.

To leverage the improved performance resulting from static scheduling, the \ompc~runtime adopted the HEFT~\cite{Topcuoglu02} scheduling algorithm to allocate tasks to worker nodes. In \ompc~runtime, the control thread creates the tasks as described before, but the worker threads are initially kept idle. When the control thread reaches the implicit barrier, it means that no more tasks will be created inside the current region and thus the entire task graph is scheduled using HEFT. After this initial scheduling phase, tasks can be dispatched to their respective worker nodes for execution, provided that all their input dependencies are already satisfied. Running this algorithm tends to cause little overhead due to its low asymptotic complexity of $\mathcal{O}(e \times p)$, where $e$ is the number of directed edges and $p$ is the number of worker nodes in the graph.

Some adaptations were made to the original formulation of the algorithm to match the programming model. First, classical tasks, the ones declared with \ttt{task} directive, are unconditionally scheduled on the head node because otherwise, we would be violating the semantics of the \omp~standard. Secondly, \ttt{target data nowait} clauses are represented as tasks in the graph, but they differ from regular \tit{target tasks} because the former does not execute any code and only represents a data transfer between processes. For this reason, these \tit{target data} tasks are always scheduled to the same process as the target task that will use or produce the data. Not scheduling both tasks in the same process would lead to data being unnecessarily sent from the producer to an intermediate process and then forwarded to the consumer process.

\section{Related Works}
\label{sec:relatedworks}

Parallelization of programs in clusters is a well-known research problem that has been extensively studied~\cite{Dean2004,Gropp1996}.

Distributed applications usually rely on the Message Passing Interface (MPI) to enable synchronization within the nodes of the cluster. While MPI implementations have proven themselves to be very efficient, the programming complexity and the lack of important features like load balancing and fault-tolerance make it complex and error-prone to use. As a result, several distributed programming frameworks have been developed like UPC++~\cite{Zheng2014}, PaRSEC~\cite{Bosilca2013}, Chapel~\cite{Chamberlain2007}, Legion~\cite{Bauer2012}, StarPU~\cite{starpu}, Charm++~\cite{Kale1993}, Fortran Coarrays~\cite{coarray}, etc. Just like \ompc, most of those distributed programming languages and libraries rely on high-performance transport layers such as MPI implementations \cite{OpenMPI2014,Huang2006}, Partitioned Global Address Space (PGAS) libraries~\cite{Bonachea2002,Nieplocha1999}, or the Unified Communication X (UCX) library~\cite{shamis2015ucx} to achieve low communication latency, portability, and scalability. However, we believe that OpenMP is easier to use than most other task programming languages, which is supported by its strong popularity. As an example, StarPU and PaRSEC are much lower-level and require programmers to explicitly write the runtime function calls. Charm++ relies on the concept of \textit{Chares} and over-decomposition which, despite being implemented as a language extension of C++, enforces a new programming model where computation is bounded to the data itself. Such characteristic forcefully splits the source code even more, making the communication patterns between Chares hard to analyze. Finally, similar observations can be made for UPC++, which, alongside its use of APGAS instead of more widespread communication models like MPI, makes it harder to be deployed in many computing environments.

Several other works have proposed the use of directive-based programming to program computer clusters and heterogeneous architecture. First, Nakao et al. proposed a new directive-based programming language similar to OpenMP, but specialized in HPC clusters~\cite{Nakao2015}. Its guidelines allow micromanaging the parallelization and communication within the distributed architecture of the cluster; its custom compiler then translates pragmas into MPI calls. The OpenACC accelerator model is used to transfer computing to a GPU within each node. If on one hand, they demonstrate very good scalability, close to a standard MPI implementation, on the other hand, the proposed extensions do not follow the OpenMP standard, are based on data-parallel programming without support for tasks, and require information about the cluster architecture (for example, the number of nodes), which reduces its portability. Duran et al. introduced another directive-based programming language, called OmpSs~\cite{Duran2011}, to program heterogeneous multi-core architectures. OmpSs was successfully used to design applications for multi-CPU (multi-socket) and multi-GPU architectures~\cite{Ozen2016}. OmpSs was also extended to support a distributed architecture~\cite{Bueno2013} based on GASNet~\cite{Bonachea2002}, a software infrastructure for PGAS languages commonly used in supercomputing. The results show scalable performance quite similar to MPI. However, the methodology for supporting distributed systems with multiple address spaces using multilevel task directives is programmatically complex. Jacob et al. introduced a new methodology based on the OpenMP accelerator model to run applications in a cluster using the MPI infrastructure~\cite{Jacob2015}: the host code is executed by the master node and the kernels transferred by the working nodes. The offloading for a set of working nodes is achieved by defining the kernel to be transferred from inside the body of a parallel loop. However, this requires several manual modifications to the application, such as programmatically splitting (and merging) the transferred data, and setting the nested DOALL loop to parallelize execution on multi-core worker nodes. Yviquel et al. investigated the use of Apache Spark~\cite{Zaharia2010} to parallelize DOALL loops of OpenMP programs. Spark is used as a cluster runtime that manages computation and communication between nodes. Although some interesting results have been demonstrated, the approach also presented some scalability limitations and lacked support for task parallelism~\cite{Yviquel2018}. Similarly, Patel et al. proposed an extension to the OpenMP accelerator model in LLVM to program remote GPU clusters using RPC on top of UCX~\cite{Patel2022}. However, their approach requires programmers to manually map the tasks to the GPUs and manually manage the data making it difficult to scale.

Unlike previous work, our approach allows simple cluster programming using fully compliant OpenMP directives. Moreover, it enables automatic load-balancing and data management by means of an underlying MPI-based runtime, which makes it highly portable and efficient on HPC clusters. We also implemented this approach in LLVM/Clang, a well-known and widely used open-source compiler infrastructure, allowing us to experiment with a set of benchmarks and real-world applications.

\section{Experiments}
\label{sec:experiments}

To evaluate OMPC and its robustness in a real-world scenario, we performed a set of experiments comparing its performance against other runtimes. Here,~\rsec{expsetup} lists the cluster environment setup as well as OMPC Bench, a supporting tool developed to improve the reproducibility and analysis of the experiments. Next, \rsec{expresults} shows and discusses the experimental results.

\subsection{Environment Setup}
\label{sec:expsetup}

The experiments to be presented in the next section were conducted at the Santos Dumont supercomputer located at the National Laboratory for Scientific Computing (LNCC) in Brazil~\cite{sdumont}. From the possible configurations, the experiments used up to 64 nodes, each with two Intel Cascade Lake Gold 6252 CPUs (24 Cores / 48 Threads) and 384 GB of RAM. The nodes are interconnected by an InfiniBand network, theoretically capable of 100Gb/s of low-latency bandwidth. Since OMPC relies on MPI underneath, MPICH (version 3.4.2) was used with its UCX backend (version 1.11.2). To further utilize the potential of the underlying network, a custom version of MPICH was compiled to enable up to 64 Virtual Communication Interfaces (VCIs)~\cite{mpivcis}, which allowed OMPC to use multiple hardware network contexts for concurrent communications. Finally, the SLURM job manager was used to manage and schedule the experiments.

Two sets of applications were used to evaluate the OMPC performance. The first, called \tit{Task Bench}, is a highly configurable synthetic benchmark designed to measure the efficiency and performance of distributed and parallel programming models~\cite{taskbench}. The second is a real-world seismic imaging geophysics application, namely \tit{Awave}, that we developed which is based on the solution of a Reverse Time Migration problem~\cite{baysal1983reverse}.

Task Bench currently has 17 implementations, including different programming models such as OpenMP and OmpSs, that do not allow distributed memory, or Swift/T and TensorFlow, that use the data flow paradigm. We chose to focus on those runtimes that have similar characteristics to OMPC (i.e., task-based paradigm over distributed systems). Therefore, the selected implementations were Charm++ and StarPU. The MPI implementation was also tested, allowing all runtimes to be evaluated against the best possible baseline.

The experiments used two main tools that enabled a high degree of reproducibility. First, we used Singularity containers for each tested runtime, providing a consistent execution environment across multiple runs, easing the validation of the results. Second, we developed OMPC Bench, a custom python tool responsible for correctly launching the experiment jobs based on a YAML configuration file. OMPC Bench was made compatible with all used runtimes, guaranteeing the same experimental parameters for all runs. Besides that, it also provides a reliable method for extracting average and dispersion statistics from multiple executions.

\subsection{Experiments Results}
\label{sec:expresults}

\begin{figure}
 	\centering
 	\includegraphics[width=1\linewidth]{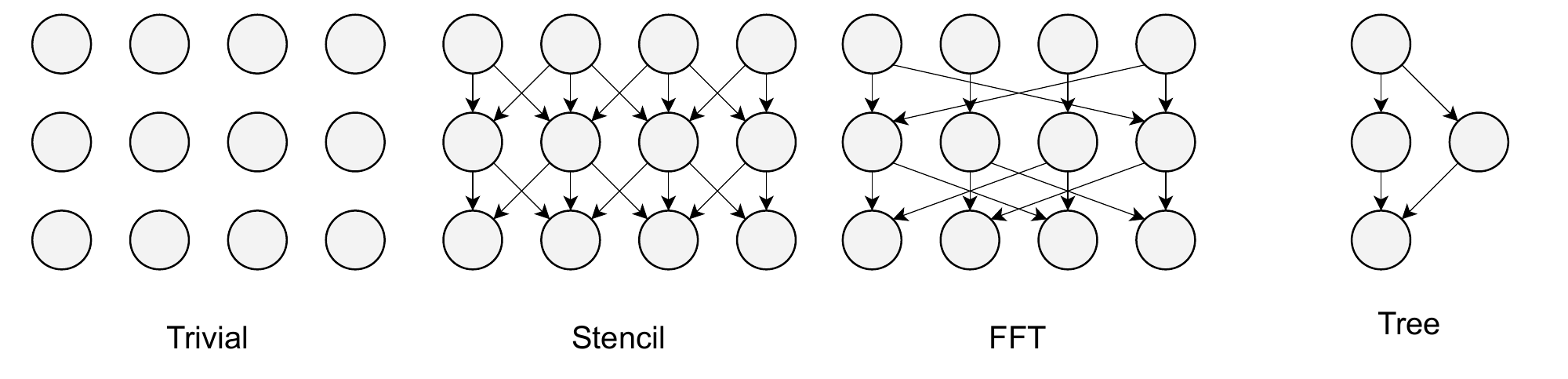}
    \caption{Task Bench dependency types used in OMPC's experiments}
 	\label{fig:deps}
\end{figure}


\begin{figure*}[!ht]
 	\centering
    \includegraphics[width=1\linewidth]{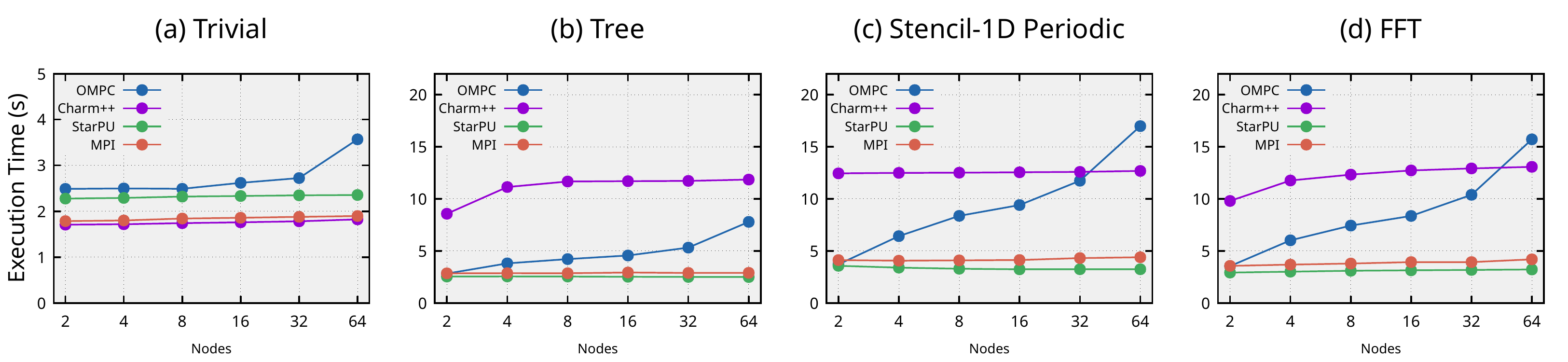}
    \caption{Execution time Scalability}
    \label{fig:speedup}
\end{figure*}

\begin{figure*}[!ht]
 	\centering
 	\includegraphics[width=1\linewidth]{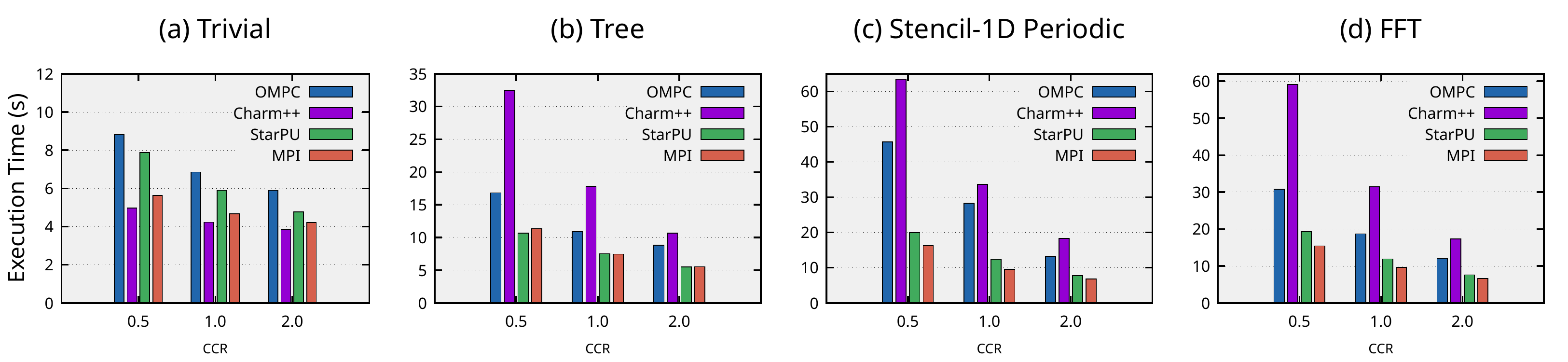}
    \caption{Execution time varying the Computation to Communication Ratio}
 	\label{fig:ccr}
\end{figure*}

\begin{figure}[!ht]
 	\centering
 	\includegraphics[width=1\linewidth]{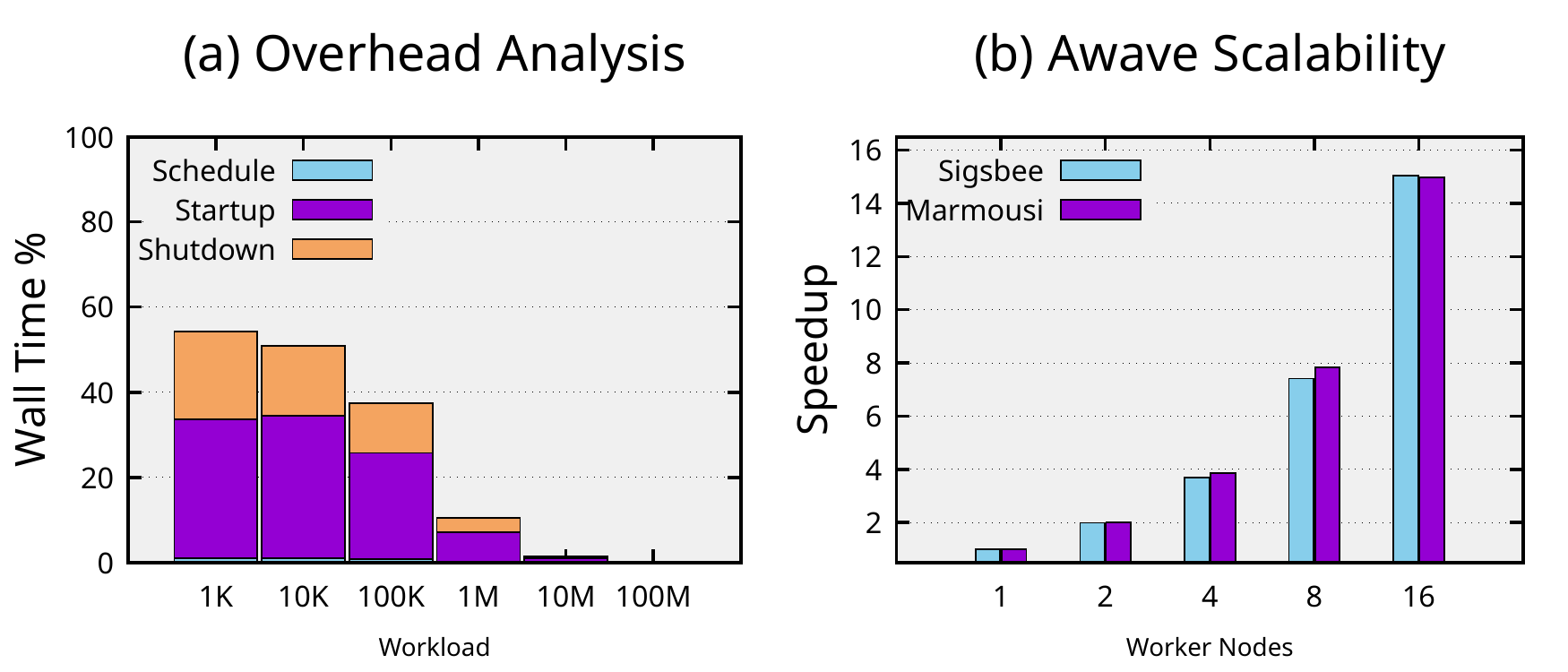}
    \caption{OMPC Runtime overhead analysis and Awave experiment}
 	\label{fig:analysis}
\end{figure}

Initially, a scalability experiment using Task Bench was performed with the goal of comparing OMPC against other runtimes (\rfig{speedup}). All experiments used $10$M iterations ($50$ms workload) per task, and distinct dependency patterns (\rfig{deps}). The x-axis shows the number of nodes ranging from 2 to 64 and on the y-axis the average execution time of 10 runs. All executions used a CCR\footnote{\textit{Computation to Communication Ratio} is a quantity that measures the ratio between the computation and the communication costs of an application.} of 1.0. The size of the task graph for each execution was defined by $2n \times 32$, where $n$ is the number of nodes. As shown in ~\rfig{speedup}, as the number of nodes increases, OMPC performs better than Charm++. OMPC achieved average speedups of 1.61x, 1.64x, and 2.43x against Charm++ using FFT, Stencil 1D, and Tree, respectively. On the other hand, it still lacks behind StarPU and MPI. The second goal of this experiment was to evaluate OMPC's weak scalability. To achieve that the task graph workload was increased at the same rate (doubling) as the number of nodes. From the graphs in ~\rfig{speedup}, it is clear that OMPC's weak scalability is strongly dependent on the communication pattern. Although the Trivial pattern somehow preserve scalability until 32 nodes, this is not the case for the Tree, FFT and Stencil-1D patterns. The results in~\rfig{speedup} show that, despite \ompc~having performance comparable to Charm++, OMPC still has some scalability issues that need to be addressed.

Next, we performed a second experiment to measure OMPC's performance against other runtimes with multiple degrees of CCR (\rfig{ccr}). For this purpose, this experiment also used Task Bench, fixing the number of nodes to $16$, the task graph to $16 \times 16$, and the execution time to $500$ms per task ($100$M iterations), but varying the amount of data exchanged between the tasks depending on the desired CCR values of $0.5$, $1.0$ and $2.0$. The experiment also used the same 4 dependency patterns as the previous one. \rfig{ccr} shows the average execution times of 10 Task Bench runs for each performance configuration. While using Tree, Stencil 1D, and FFT dependencies, OMPC either matched or surpassed Charm++'s performance, showing average speedups of $1.53$x, $1.34$x, and $1.41$x, respectively. Again, although not far behind, OMPC still struggles against StarPU and MPI in all scenarios. Interestingly enough, OMPC handled well the many CCR values, showing similar variability as StarPU and MPI. This cannot be said about Charm++, which had its performance dramatically decreased when the communication took most of the execution time of the benchmark.

The goal of the third experiment (\rfig{analysis}(a)) was to characterize the overhead of the OMPC runtime when executing Task Bench graphs with different task sizes (durations). The baseline selected for the experiment has: (a) 1 head node and 1 single worker node; and (b) a $1\times16$ Task Bench graph using the \tit{Trivial} dependency pattern that does not have dependencies between tasks. The rationale for this baseline is to force tasks to run on a single node and with a single thread to assess startup and shutdown time, and measure how much time is spent on control versus how much time is spent on task execution. The task workload varies from 1K iterations, i.e., the number of iterations of the internal loop of the task (about $0.02$ms), to $100$M iterations ($500$ms). For the correct interpretation of the results, the following definitions were used: \tit{Startup} overhead represents the time elapsed from the beginning of the process to the creation of the gate thread; \tit{Shutdown} overhead represents the time elapsed from the destruction of the gate thread to the end of the process; and, \tit{Scheduling} overhead measures how long it took to schedule the entire graph. All the times are normalized with respect to the total execution time (Wall Time). Based on the graph of \rfig{analysis}(a) one can observe that the initialization and shutdown times seem to be constant in all invocations. All invocations have an interval of approximately 4.7ms at the main node just after the first event. The overhead goes down quickly as the tasks get bigger. $10$ms per task seems like a reasonable lower bound to get a small overhead ($<25$\%). The constant overhead for the runtime fluctuates around $25$ms. The conclusion is that the OMPC runtime shows negligible overhead with large tasks ($> 50$ms per task, $> 10$M iterations). With smaller tasks ($<5$ms per task, $<1$M iterations), the runtime overhead is dominant but these tasks are very fine-grained relative to real applications in the HPC world.


Seismic imaging is a problem from the Geophysics domain that is central to oil prospecting. It consists of generating a model of the subsurface from seismic data collected on the surface. To achieve such results a couple of techniques can be employed: Reverse Time Migration (RTM)~\cite{baysal1983reverse}; Common-Reflection Surface (CRS); and Full-Waveform Inversion (FWI). Awave is an implementation of the RTM algorithm that consists of numerically solving the acoustic wave equation using the finite differences method. At each time step, one pressure field is computed for the down-going wave and another for the up-going wave. After all time steps are computed, the intermediate fields are correlated to produce the final image. This process is repeated many times for each seismic trace collected, also referred to as a \tit{shot}. Shots can be computed independently from each other and have their results combined, essentially forming a sequence of images. In this experiment, a single shot is assigned to each worker node. To measure application performance, we used two known 2D models: (1) \tit{Sigsbee}~\cite{soubaras2007velocity}, a finite-difference acoustic dataset with constant density, and (2) \tit{Marmousi}~\cite{brougois1990marmousi}, a complex synthetic structural model with strong horizontal and vertical velocity changes. \rfig{analysis}(b) shows that weak scalability can be achieved when running Awave with OMPC as the number of nodes increases up to 16. For both 2D models performance remains close to the ideal speedup, highlighting the capability of the \ompc{} runtime. This can be explained by the fact that Awave tasks have a much higher granularity than Task Bench ones.

\section{Future Work}
\label{sec:futurework}
OMPC has for goal to ease the burden of developing distributed applications with a single programming model, seamlessly distributing tasks over cluster nodes. Unfortunately, some limitations are still present, which may decrease performance and impose some restrictions on the OpenMP utilization, hurting clarity. This section explores some of these limitations, hinting at possible solutions that may be deployed in the future.

Although OMPC has better performance than Charm++, the scalability tests clearly show it may not scale well with a higher number of cluster nodes. In fact, considering how libomptarget implements the "nowait" clauses, such performance deficiency may come from the increase in concurrently executed tasks instead of the increase in the number of nodes itself. In LLVM, an OpenMP thread at the head node is always blocked, waiting for a target region to complete (even when it is marked as \ttt{nowait}). This means that we can have as many in-flight tasks  as we have threads on the head node, decreasing the performance on the scalability tests results with 32 and 64 nodes when considering the used cluster configuration. Such restriction is not present in the other Task Bench implementations. Integrating the libomptarget library and the OpenMP runtime more asynchronously could resolve the problem. Here, the execution of a \ttt{target nowait} region could be done in two steps. First, all operations needed to execute the task could be sequentially and asynchronously dispatched and added to a \tit{target context handle} (e.g., operation queue) present in the task structures inside the OpenMP runtime. Then, as a second step, the bounded OpenMP task could be re-scheduled for execution, allowing the context handle to be checked for completion in the future, re-scheduling it again if there are still pending operations. Such a mechanism would allow the threads to dispatch many target regions concurrently, even letting a single OpenMP thread manage an "infinite" number of target regions, thus resolving the problem not only for OMPC but for all target devices. In fact, this limitation has already been pointed out by the libomptarget developers~\cite{Tian2022}, but has not been entirely fixed yet.

Of course, the centralized nature of the runtime around a head node is also a large problem and may become even more apparent with a higher node count. In OMPC the amount of concurrent work computed by the cluster is limited by how fast the head node can dispatch it. A more distributed runtime, where the worker nodes take some of the burdens away from the head node (e.g., proactively forwarding data and resolving task dependencies without any head node), may solve the problem. But, currently, this is a lesser issue when compared to the synchronicity of OpenMP's integration with libomptarget.

The OMPC runtime is not an ordinary OpenMP offload target. The standard treats each device individually, meaning the programmer has to specify where a memory region or a target block will be sent/executed. On the other hand, OMPC sees all the cluster nodes as individual devices in a single device group, automatically mapping memory and target regions to the whole cluster. Thus, some extensions and limitations were imposed on how a program uses the OpenMP target clauses. Currently, there is no way of pinning data or target regions to a specific cluster node, meaning the programmer cannot manually optimize the data distribution to a defined environment. This cannot be mitigated without extending the OpenMP standard to support the aggregation of multiple instances of the same device type under a unique ID, a device group. With such extension in hands, an OMPC user could map different target clauses to specific cluster nodes or the whole cluster, leaving the data coherency to the runtime's data management system.

Regarding the automatic data distribution, the program must always include the memory buffers that will be written to in a target region as an out-dependency, meaning that it cannot use artificial/dummy data dependencies to order the execution of tasks, as it would be normally possible with OpenMP code. A possible solution for this problem would be to implement a different write detection mechanism based on page write protection, where the runtime marks a mapped data region as dirty by intercepting any write operations to its allocated pages. Since the runtime would place each device allocation inside a single memory page and it would intercept memory operations by OS interruptions, some performance, and space analysis would be required to evaluate its applicability.

Regarding other offloading devices, the libomptarget implementation also incurs another limitation on how OMPC can be used: currently there is no support for a second-level offloading. To utilize an accelerator inside a node, a program needs to use another library/programming model (e.g., CUDA, OpenCL). We hope that work can be done in the future to lift such restrictions, at least for x86 offloading devices like OMPC, allowing OpenMP directives to be used for cluster nodes distribution, and local accelerator programming using nested target regions.

Last but not least, there are currently no optimizations regarding one-to-many data transfers among the nodes in a proactive manner. All communications are done in a peer-to-peer fashion and require the orchestration of the head node (even though the data transfer does not use it as a middle man), which is not optimal. We are currently working to automatically detect such communication cases using the task graph itself, implementing a broadcast event that can distribute the data to many nodes without any intervention from the head node at each communication.

\FloatBarrier

\section{Conclusion}
\label{sec:conclusion}

This paper presents OpenMP Cluster (OMPC), a task-parallel programming model and runtime that allows applications to target computer clusters. It extends LLVM's OpenMP implementation by introducing a new offloading target, making its adoption easy by using already familiar parallel programming concepts. It is highly compatible with many HPC clusters by utilizing MPI as its communication layer and containers to ensure a well-defined execution environment. Since there is no need for integration with new libraries and to learn entirely new languages, OMPC also becomes a great option to adapt existing projects to distributed systems.

We showed through the synthetic benchmark, Task Bench, that OMPC can be competitive with other distributed runtimes. On CCR tests, OMPC demonstrated a speedup of $1.41$x, $1.34$x, and $1.53$x against Charm++ using FFT, Stencil 1D, and Tree, respectively. OMPC also surpasses the competing runtimes on most of the scalability tests, with average speedups of $1.61$x, $1.64$x, and $2.43$x for the same dependency types, although the pattern only held up until 32 computer nodes. We also demonstrated that OMPC can scale on Awave, a real seismic imaging application, approaching near-linear scaling. Other Task Bench runtimes still performed better than OMPC, with the low-level MPI implementation achieving 1.4x to 2.9x better performance, an expected result considering that the application can greatly tailor its communication patterns and better distribute the program execution.

OMPC is still an evolving project with many performance and ease of use improvements already planned. The OMPC's intrinsic overhead alongside the lack of better integration between the libomptarget library and OpenMP tasks were demonstrated to be the real culprit in decreasing the runtimes performance. With such limitations solved, OMPC can become an effective programming model for HPC systems and accelerators.

\section*{Acknowledgments}
This work is supported by Petrobras under grant 2018/00347-4, by the Center for Computing in Engineering and Sciences (CCES) and São Paulo Research Foundation (FAPESP) under grant 2020/08475-1, and by the National Council for Scientific and Technological Development (CNPq) under grant 402467/2021-3. We would also like to thank the LNCC for granting the computational resources at the Santos Dumont supercomputer that were crucial to perform the experiments for this project.

\bibliographystyle{acm}
\bibliography{biblio-herve,biblio-guido,biblio-outros}

\clearpage

\end{document}